\documentclass{jpsj3}
\usepackage{txfonts}

\usepackage{color}
\usepackage{graphicx}
\usepackage{dcolumn}
\usepackage{bm}
\usepackage[normalem]{ulem} 

\newcommand{\al}{a_l}

\newcommand{\Rom}[1]{\MakeUppercase{\romannumeral #1}}

\title{
  Universal and Non-Universal Correction Terms of Bose Gases
  in Dilute Region: a Quantum Monte Carlo Study
}

\author{
  Akiko Masaki-Kato$^{1,2}$,
  Yuichi Motoyama$^3$ and
  Naoki Kawashima$^{3,4}$\thanks{kawashima@issp.u-tokyo.ac.jp}
}
\inst{
  $^1$Computational Condensed Matter Physics Laboratory, RIKEN, Wako,
  Saitama 351-0198, Japan \\
  $^2$Computational Materials Science Research Team,
  RIKEN Center for Computational Science (R-CCS),  Kobe, Hyogo 650-0047, Japan \\
  $^3$The Institute for Solid State Physics, The University of Tokyo, Chiba 277-8581, Japan \\
  $^4$Trans-scale Quantum Science Institute, The University of Tokyo, Bunkyo, Tokyo 113-0033 Japan
} 

\abst{
  We study dilute gases of interacting bosons at zero-temperature in the region where the system is characterized only by the $s$-wave scattering length. We carry out quantum Monte Carlo simulation of the Bose Hubbard model and a continuous-space hard-core model. Fitting the extended Lee-Huang-Yang formula to the Monte Carlo results establishes the detailed mapping from the lattice model to the continuous field theory characterized only by the $s$-wave scattering length. Our estimate of the intrinsic $s$-wave scattering length $a_s$ of the Bose-Hubbard model is $a_s/\al = 0.316(2)$ 
  where $\al$ is the lattice constant. It turned out that inclusion of the universal second correction term of $O(n \log n)$ makes the fitting worse while it is restored by inclusion of the non-universal third correction term, of which the existence was predicted analytically.
}

\begin{document}
\maketitle

\section{Introduction}

Ground-state properties of homogeneous dilute bosons have been well studied since Bogoliubov theory was established\cite{Bogoliubov}.
It is widely believed that interacting bosons show universal behavior 
in the dilute limit, i.e., their thermodynamic properties do not depend on the
details of the interaction but are characterized only by
the number density $n$ and the s-wave scattering length $a_s$.
Specifically, for the ground-state energy as a function of the gas parameter
$\sqrt{n{a_s}^3}$,
Lee, Huang and Yang obtained the first-order correction to the Bogoliubov
mean-field approximation (LHY correction) using the perturbation
theory\cite{Lee, Fetter}, which was followed
by higher-order terms obtained by Wu\cite{Wu},
\begin{eqnarray}
  \frac{E}{N}
  &=& \frac{2\pi\hbar^2na_s}{m}\left[ 1+
    \left(\frac{128}{15\sqrt{\pi}} \right)\sqrt{na_s^3}\right.\nonumber\\
  &+&
    \left.\frac{8(4\pi-3\sqrt{3})}{3}na_s^3\log(na_s^3)+ O(na_s^3) \right],
\label{eq:E_theory}
\end{eqnarray}
where $a_s$, $n$ and $m$ are the $s$-wave scattering length, the number density
and the mass of a boson, respectively.
The condensate density $n_0$
was obtained by Bogoliubov\cite{Bogoliubov} as
\begin{eqnarray}
n_0 = 1-\frac{8}{3\sqrt{\pi}}\sqrt{na_s^3}.
\label{eq:C}
\end{eqnarray}
It was shown \cite{Brueckner,Beliaev,Hugenholtz,Lieb}
that the first three terms (up to the one with the logarithmic factor)
do not depend on the character of the two-body interaction, whereas
the higher terms do.
Below we call the first three terms in (\ref{eq:E_theory}) ``universal'' and
the low-density region in which the energy is well approximated by the
first three terms the universal region.
While we are mainly interested in the universal behaviors,
in what follows, we study a broader range that seems to
require non-universal terms for a full explanation.

Recent developments in low-temperature atomic physics
shed light on the fundamental properties of bosonic systems. The static properties of bosonic systems have been actively investigated by experiments and theories on ultra-cold atoms over the past two decades. Numerical simulations are effective for accurately treating inter-atomic interactions with various types of potentials and with a wide range of strength to corroborate the analytical solutions.
For the continuous-space systems, finite-temperature quantum Monte Carlo studies
have provided estimates of the critical temperature as a function of $na_s^3$ for the 3D hard-sphere gas in the homogeneous space\cite{Gruter} and with the external parabolic potential\cite{Pearson}.
Using the diffusion Monte Carlo (DMC) method at $T=0$ Giorgini $et$~$al$
\cite{Giorgini} presented a detailed comparison of the numerical results with the LHY corrections to the Bogoliubov mean-field.
They obtained results in the universal region while
they also found that the logarithmic correction by Wu\cite{Wu} is
insufficient for predicting experimental results for cold bosons
and liquid $^4$He with realistic densities\cite{Giorgini}.
This may be because the particle density in the experiments was too high for the perturbative expansion Eq.(\ref{eq:E_theory}) to be a quantitatively good approximation.

As mentioned above, Bose gas with the delta-function potential in continuous space and the Bose-Hubbard model on the cubic lattice are the two most fundamental models for bosons, and each of them has been studied extensively for many years.
While it is obvious from the theoretical point of view that the two models are essentially equivalent in the diluted limit, where the short-range details
of the system are irrelevant, the quantitative correspondence between them has never been studied in detail. Once the quantitative correspondence has been established, one will be able to benefit equally well from the study of either one of the models.
In the universal region, we expect that the behaviors of the two models
are characterized only by the s-wave scattering length. In the present short paper, by studying the dilute region of Bose gases using the quantum Monte Carlo (QMC) methods for lattice and continuous spaces, we establish the quantitative correspondence between the two, and verify the validity of the analytical low-density expansion by Lee, Huan and Yang and other groups. In particular, we obtain a concrete estimate of the s-wave scattering length $a_s$, in the unit of the lattice constant, $\al$.

\section{Lattice and Continuous Models}
For the continuous system, we consider the Hamiltonian of interacting bosons
expressed as
\begin{equation}
  {\cal H}_c=-\frac{\hbar^2}{2m} \sum_i \frac{\mathrm{d}^2}{\mathrm{d}{\bf r}_i^2}
  + \sum_{i<j} V({\bf r}_i,{\bf r}_j),
\label{eq:cont}
\end{equation}
where ${\bf r}_i$ is the spatial coordinate of an $i$-th particle and $V$ is a
two-body interaction.
For the lattice system, we employ the hard-core Bose-Hubbard model in the cubic
lattice given by
\begin{eqnarray}
  {\cal H}_l=-t\sum_{\langle i,j\rangle}b^\dagger_ib_j,
\label{eq:BH}
\end{eqnarray}
where $t$ is the hopping constant, and $b_i (b_i^\dagger)$ the annihilation (creation) operator. We restrict the Hilbert space to ensure the hard-core condition $n_i = 0,1$, where $n_i$ is the number operator at $i$th site. In our calculation of the lattice systems, we introduce the chemical potential and simulate the grand-canonical ensemble. However, to make the comparison with previous fixed-number\cite{Giorgini} calculations more accurate, we extract the canonical-ensemble averages from the output of our grand-canonical simulation. Therefore, the numerical results presented below are the canonical-ensemble averages.

\begin{figure}[ht]
  \includegraphics[angle=0,width=8.5cm,trim= 0 0 0 0,clip]{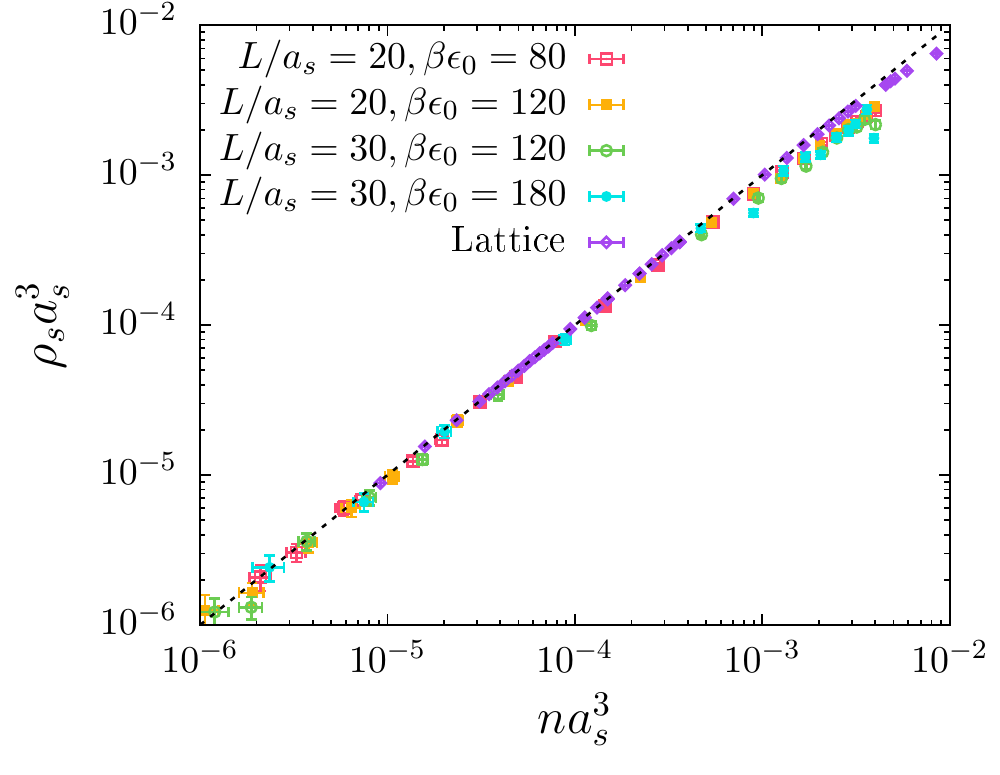}
  \caption{
    (Color) The dimensionless superfluid density $\rho_s a_s^3$ of the continuous and the lattice models plotted against the density. The grand-canonical ensemble averages are plotted. Purple diamonds represent $\rho_s$ of the lattice system with $L/\al=32, \beta t=64$ and $\gamma=0.3151$. (See Table I) Other points are for the continuous-space system with $L/a_s=20, 30$ and $\beta\epsilon_0=80 - 180$ respectively. The dashed line corresponds to the ``perfect'' superfluid ($\rho_s/n=1$).
  }
  \label{fig:wn}
\end{figure}

At first, we show the comparison between the superfluid densities of the continuous-space system and the lattice system.
The superfluid density is defined as the free-energy response to the infinitesimal flux.
As is well known, the superfluid density can be expressed with the winding number of worldlines with the periodic boundary condition\cite{Pollock} as, 
$\rho_s=\frac{m}{\hbar^2}\frac{\langle|{\bf W}|^2\rangle}{d\beta L^{d-2}}$.
To study the connection between the different systems, we examine the dimensionless superfluid density
\begin{equation}
\rho_s a_s^3=\frac{\langle|{\bf W}|^2\rangle}{2d\beta\epsilon_0\tilde{L}},
\end{equation}
where $d$ is the space dimension, $\tilde{L}\equiv L/a_s$ is the dimensionless linear size of the system. For the unit of the energy, we adopt
$\epsilon_0 \equiv \frac{\hbar^2}{2ma_s^2}$.
Here we introduce $\gamma$ as the $s$-wave scattering length of the lattice system measured in the unit of the lattice constant, i.e., $\gamma \equiv a_s/\al$. The s-wave scattering length is the key-parameter because it is the only parameter that characterizes the system in the universal low-density region. While its value is obvious in the continuous-space delta-function model, it is not obvious for the Bose Hubbard model. In fact, to our knowledge, its value has never been estimated before. In order to establish the quantitative correspondence between the continuous-space and the discrete-space systems, it is the key quantity.
The hopping constant $t$ for the hyper-cubic lattice is related to $m$ by $t=\frac{\hbar^2}{2m\al^2}$. As an order parameter of the superfluid state, $\rho_s$ is more convenient in the simulation methods we use
for the present work than $N_0$, which is an off-diagonal quantity. For continuous-space model (Eq. (\ref{eq:cont})), we adopt the hard-core interaction
\begin{equation}
  V({\bf r}, {\bf r}') =
  \begin{cases}
    \infty & \text{if} \quad |{\bf r} - {\bf r}'| \le r_\text{c} \\
    0 & \text{otherwise}
  \end{cases},
\end{equation}
with the hard-core radius $r_\text{c}$,
which is the same as the $s$-wave scattering length $a_s$. 
the world-line Monte Carlo (WLMC) method based on
the continuous-time directed-loop algorithm
for the lattice system\cite{Kawashima,Kato,DSQSS},
and WLMC method with worm update
for the continuous-space model~\cite{Ceperley, Boninsegni}.
For the discretization step of the imaginary time in simulations of the continuous-space model,
we have used $\Delta \tau = 0.05\epsilon_0^{-1}$, 
which we find small enough to ensure the convergence of the results. Points for the continuous system in Fig.~\ref{fig:wn} represent
particle density dependence of the superfluid density for several combinations of system size and inverse temperature, i.e., $L/a_s = 20, 30$ and $\beta\epsilon_0 = 4L/a_s, 6L/a_s$. In the scale shown in Fig.~\ref{fig:wn}, the size and the inverse-temperature dependence is hardly recognizable.
By using $\gamma=0.3151$ (See Table I),
obtained by the fitting based on the Bayesian regression as mentioned below, the lattice results and the continuous-space results agree with each other in the low-density region $na_s^3 \lesssim 10^{-4}$. In this region, the system is a nearly-perfect condensate as is indicated by the dashed line, i.e., almost all particles are in the superfluid component.

\section{LHY Formula Fitting to Lattice Model}
  Next, we consider the connection between 
the $s$-wave scattering length and the lattice constant. Let us consider the ground-state energy per particle in the continuous space and in the discrete space. To compare the two systems on equal footing, we define the dimensionless energy per particle $\tilde{\epsilon} \equiv N^{-1} \langle {\cal H} \rangle/\epsilon_0$. We can express this quantity by the universal dimensionless function:
\begin{equation}
    \tilde{\epsilon}= f(na_s^3),
    \label{eq:E_form}
\end{equation}
where $n=N/\Omega$ is the particle density.
By comparing the energy densities of the two systems we can estimate the unknown parameter $\gamma$. In this study we assume a form that includes the analytical prediction as a special case. Specifically, we consider the following form for $f$ in (\ref{eq:E_form}).
\begin{eqnarray}
    f(x) & = &
    4 \pi x \times
    \left(
    1 + c_1 \frac{128}{15\sqrt{\pi}} \sqrt{x}
    \vphantom{+ c_2 \frac{8(4\pi-3\sqrt{3})}{3} x\ln x}
    \right.
    \nonumber \\
    & &
    \left.
    \qquad
    + c_2 \frac{8(4\pi-3\sqrt{3})}{3} x\ln x
    + c_3 x
    \right),
    \label{eq:NLF}
\end{eqnarray}
In (\ref{eq:NLF}), the analytical predictions are $c_1=1$\cite{Lee} and $c_2=1$ \cite{Wu}.
It was argued that $c_1$ and $c_2$ are universal, independent of the shape of the model's potential, whereas $c_3$ is not\cite{Wu,Hugenholtz}.

Because our calculation simulates the grand-canonical ensemble, we can obtain canonical-ensemble averages for many different values of $n$ by a single Monte-Carlo run.
For this conversion, we accumulate measured energy (and other quantities as well) separately for each total particle number during our grand-canonical Monte Carlo simulation. The separate accumulation makes it possible to compute the canonical
averages of the energy as a function of the density in a narrow but finite range centered around the grand canonical average of the density.

In our simulation, the range of the system size is from $L=16$ up to $64$, and the inverse temperature $\beta=32$--$256$. Although the data for the largest system sizes are shown in Fig.~2 together with the data for smaller systems, we could not collect a sufficiently large number of independent samples that could contribute to the minute fitting discussed below, and hence they are not used in the fittings that require precision. Only the system of the size $(L,\beta) = (32,64),(32,96)$ and $(48,96)$ are used in the other figures and Table I. For each value of the chemical potential, we have confirmed that the numerical estimates have reached the saturation as a function of $L$ and $\beta$ within the statistical error, i.e., the results shown in all the figures are for sufficiently large $L$ and $\beta$ so that they can be regarded as those of the thermodynamic limit at zero temperature in the scale adopted for each figure.

Together with the estimated correction to the Bogoliubov mean-field values, we also show the fitting function Eq.~(\ref{eq:NLF}). In particular, we have explored the dilute region where the systems' behaviors do not depend on the specific form of the inter-particle interaction nor the discreteness of the space, apart from non-universal corrections such as the $c_3$ term mentioned
above. In the WLMC simulation for the lattice system, the total energy can be measured through the number of kinks $N_k$ in world lines:
\begin{equation}
  \langle {\cal H}_l \rangle = - N_k / \beta.
  \label{eq:Nk}
\end{equation}
\begin{figure}[ht]
  \includegraphics[angle=0,width=8.5cm,trim= 0 0 0 0,clip]{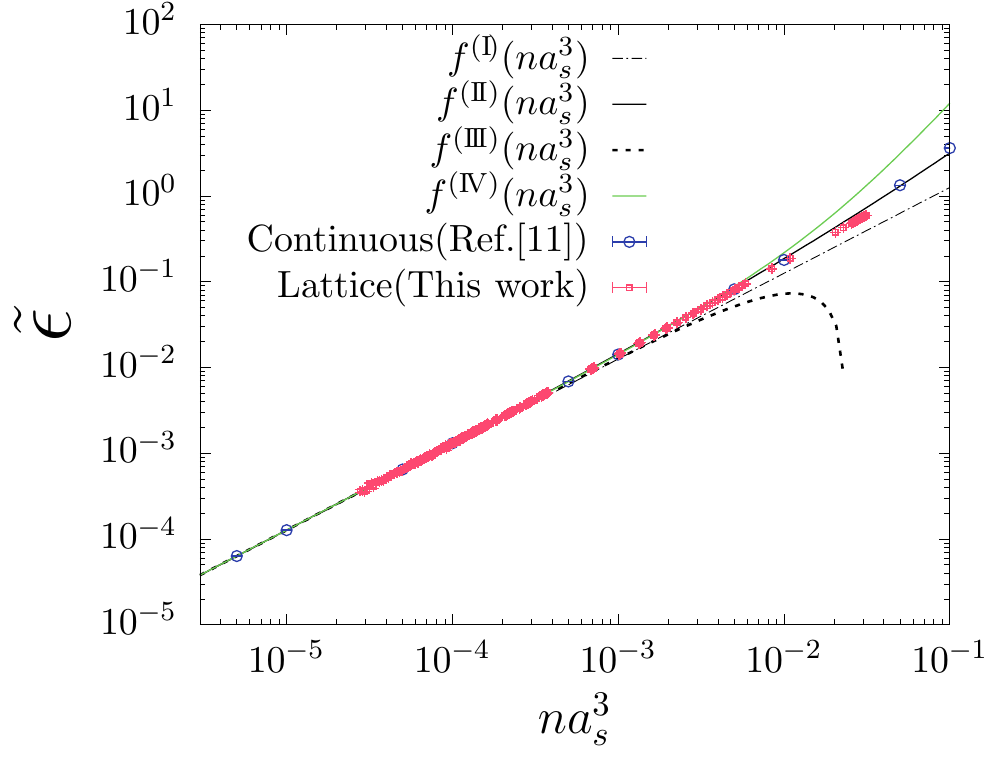}
  \caption{
    (Color online) The canonical ensemble average of the dimensionless energy per particle of the Bose-Hubbard model in three dimensions as a function of the parameter $na_s^3$. The errors are shown but too small to be visible. The fitting curves are obtained from (\ref{eq:NLF}) by setting the coefficients in Table I, except for $\gamma$. In all cases, the s-wave scattering length was set $\gamma\equiv a_s/\al=0.32$, which is not necessarily
    the optimal value for each case in Table I,
    although the deviation from the optimal fitting
    is negligible in the shown scale.
  }
  \label{fig:EN}
\end{figure}
Figure~\ref{fig:EN} shows $\tilde{\epsilon}$ as a function of $na_s^3$. We can see that our WLMC results roughly agree with the previous DMC results of the continuous-space model up to the density around $n a_s^3 \sim 0.003$, while the deviation becomes detectable at larger density
$n a_s^3 \sim 0.001$, which is consistent with the previous observation\cite{Giorgini}.

\begin{figure}[ht]
\includegraphics[angle=0,width=8.5cm,trim= 0 0 0 0,clip]{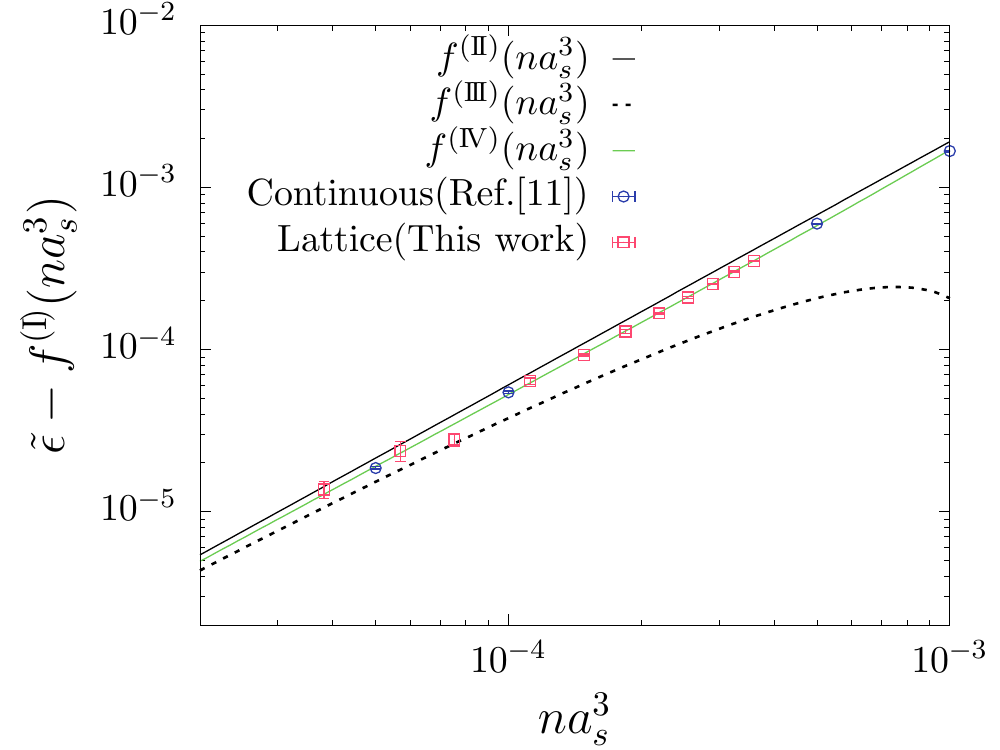}
\caption{
  (Color online) The interpolated canonical ensemble average of the correction to the Bogoliubov mean-field value of the scaled energy, $\tilde{\epsilon}-4\pi n a_s^3$, as a function of the dimensionless parameter $na_s^3$. We use $f(x)-4\pi x$ with $f$ in Eq~(\ref{eq:NLF}) for the fitting function
  with $\gamma$ and $c_3$ as the fitting parameters while $c_1=c_2$ are fixed to be 1. The estimates of these parameters are $\gamma=0.3151$ and $c_3=119$ optimized for $f^{\rm (\Rom{4})}$.
}
\label{fig:fitting}
\end{figure}
\begin{table}[hbtp]
  \centering
  \begin{tabular*}{8.5cm}{@{\extracolsep{\fill}}clccc}
    \hline
    \ fit. func.   & \quad $\gamma$   &    $c_1$  &  $c_2$    & $c_3$ \\
    \hline \hline
    \Rom1   & 0.33709(6) & 0         & 0         & 0      \\
    \Rom2   & 0.31178(5) & 1         & 0         & 0      \\
    \Rom3   & 0.32574(7) & 1         & 1         & 0      \\
    \Rom4   & 0.3151(2)  & 1         & 1         & 119(3) \\
    \hline
  \end{tabular*}
  \caption{
    Values of the parameter $\gamma$ and the coefficients
    used in drawing curves shown in the figures.
    They are estimated using the canonical ensemble data
    from the system sizes $(L,\beta) = (32,64),(32,96)$ and $(48,96)$.
  }
  \label{table:c}
\end{table}
In order to see the differences among various fittings more clearly, we show in Fig.~\ref{fig:fitting} the result of $\tilde{\epsilon}$ with the leading term (the Bogoliubov mean-field value) subtracted. We note here that, for this finer comparison, the original canonical ensemble averages are too noisy, because, as we stated above, we have split a grand-canonical ensemble simulation into many smaller pieces of canonical ensemble simulation, which makes the number of statistically independent data contributing each canonical average small. Therefore, for finer comparison such as Figures \ref{fig:fitting} and \ref{fig:wnII}, we apply the linear fitting to the resulting canonical averages (coming from the same simulation at the fixed chemical potential) as a function of the density. Based on the interpolation using this fitting, we obtained for each grand-canonical ensemble simulation the interpolated canonical average of the energy at the central value of the density. However, because the system sizes and the inverse temperatures are sufficiently large, the difference between the canonical ensemble and the grand-canonical ensemble is not significant in any of the figures.

There are three curves in Fig.~\ref{fig:fitting} corresponding to different places of truncation of Eq.~(\ref{eq:NLF}). While the parameters used for each curve and their numerical estimates are obtained by optimizing the fitting to the (un-interpolated) canonical ensemble data, which have a one-to-one correspondence with the interpolated canonical data shown in Fig.~\ref{fig:fitting}. The actual figures thus obtained are summarized in Table~\ref{table:c}.
In the case \Rom{1}, \Rom{2} and \Rom{3}, the relative $s$-wave scattering length $\gamma$ is the only fitting parameter whereas, for \Rom{4}, $\gamma$ and $c_3$ are optimized. The optimal values and the associated statistical errors
of the parameters were obtained by the grid search.
The functional form \Rom{2} with only the first universal correction seems to explain the numerical data rather well, with a slight but statistically
significant deviation as we see in Fig.\ref{fig:fitting}.
The curve \Rom{3} shows that the inclusion of the universal $c_2$ term makes the whole fitting worse.
The reasonable fitting in a broad range of the density is recovered by including the $c_3$ term (\Rom{4}). This term was predicted analytically as a non-universal correction term that may depend on the short-range physics\cite{Wu, Sawada, Hugenholtz}. However, the numerical estimate of its amplitude has been missing. The fitting \Rom{4} has the broadest range of good fitting as we see
in Fig.\ref{fig:fitting}. A more detailed comparison between the different functional forms
can be found in Appendix. 

While \Rom{4} is the most consistent with the analytical predictions, the estimates of Table I are based on the data from different $L$ and $\beta$, assuming that the size dependence can be negligible relative to the statistical error of the Monte Carlo simulation. Although we have observed no apparent inconsistency in this assumption, to obtain a more reliable estimate, we fit the functional form \Rom{4} to the data from different system sizes separately. As a result we have obtained $(\gamma,c_3) = (0.3151(3),116(4)), (0.315(1),130(40))$ and $(0.316(2),125(5))$, for $(L,\beta) = (32,64), (32,96)$ and $(48,96)$, respectively. We did not use data from the larger systems because they are too noisy and showed no significant difference from the estimate quoted above beyond the statistical error. From these results, we quote the following estimate that covers all the three mentioned above, as our conservative but reliable estimate of the intrinsic $s$-wave scattering length and the coefficient $c_3$ of the Bose Hubbard model:
\begin{equation}
  a_s/\al = 0.316(2) \quad\mbox{and}\quad c_3 = 130(40).
  \label{eq:TheEstimate}
\end{equation}

\section{Summary and Remarks}
In summary, we carried out quantum Monte Carlo simulations of the Bose-Hubbard model on the square lattice and the continuous-space hard-core Bose gas model. By comparing the numerical results with the analytical prediction of the LHY formula with additional non-universal correction, we determine two unknown parameters, i.e., the $s$-wave scattering length and the amplitude of the non-universal correction, thereby establishing the quantitative mapping from the lattice model to the
field-theoretical model characterized only by the $s$-save scattering length. Our estimate for the intrinsic $s$-wave scattering length 
is $0.316(2)$ in the unit of the lattice constant.

We note here that there is an analytical prediction
that the amplitude of the universal second correction is different\cite{Tsutsui2013,Tsutsui2016} from the value used in this paper. The prediction was made based on the formalism discussed previously\cite{Kita2009}. However, this formalism was criticized recently\cite{Watabe2020} for its multiple counting of the diagram of the single-particle Green's function. Since the independent and precise numerical estimation of the coefficients $c_1$ and $c_2$ in (\ref{eq:NLF}) requires better accuracy and computation of much larger systems, we have to leave it for a subject of the future work. Until then, our estimates of the parameters should be regarded as the ones obtained with the assumption that the analytical form (\ref{eq:NLF}) is correct (with $c_1=c_2=1$).

The present method for obtaining model parameters by comparison between the lattice model and the continuous model can be used in the estimation of the strength of the tight-binding hopping not only in dilute cases but also in more general cases.
For example, simulations of liquid $^4$He absorbed on the graphite\cite{Nakamura} attract attention, especially in the second layer where the existence of novel quantum phase has been discussed\cite{Nakamura}. However, it is not easy to obtain complete convergence with respect to system size, inverse temperature and equilibration, in simulations of continuous systems\cite{Corboz}.
While it is not trivial to construct a lattice model that explains the realistic Helium system with high accuracy, the present work may be useful in doing so. Since parallelization technique has been developed for lattice systems\cite{Masaki}, this is a promising direction for future works.

\begin{acknowledgment}
  This work was supported in part by Grants-in-Aid for Scientific Research No.17K14361 and 18H01183.
  We are thankful for providing computational resources of the K computer in RIKEN R-CCS through the HPCI System Research Project (Project Nos.: hp160152, hp170213, hp180098, hp180129 and hp180225). The computations were performed also on supercomputers at Advanced Center
  for Computing and Communications (ISC), RIKEN, and at Supercomputer Center, ISSP, the
  University of Tokyo. NK was supported by ImPACT Program of Council for Science, Technology and Innovation (Cabinet Office, Government of Japan).
\end{acknowledgment}


\appendix
\section{More detailed comparison}
\begin{figure}[ht]
  \begin{center}
    \includegraphics[angle=90,width=7cm,trim= 0 0 0 0,clip]{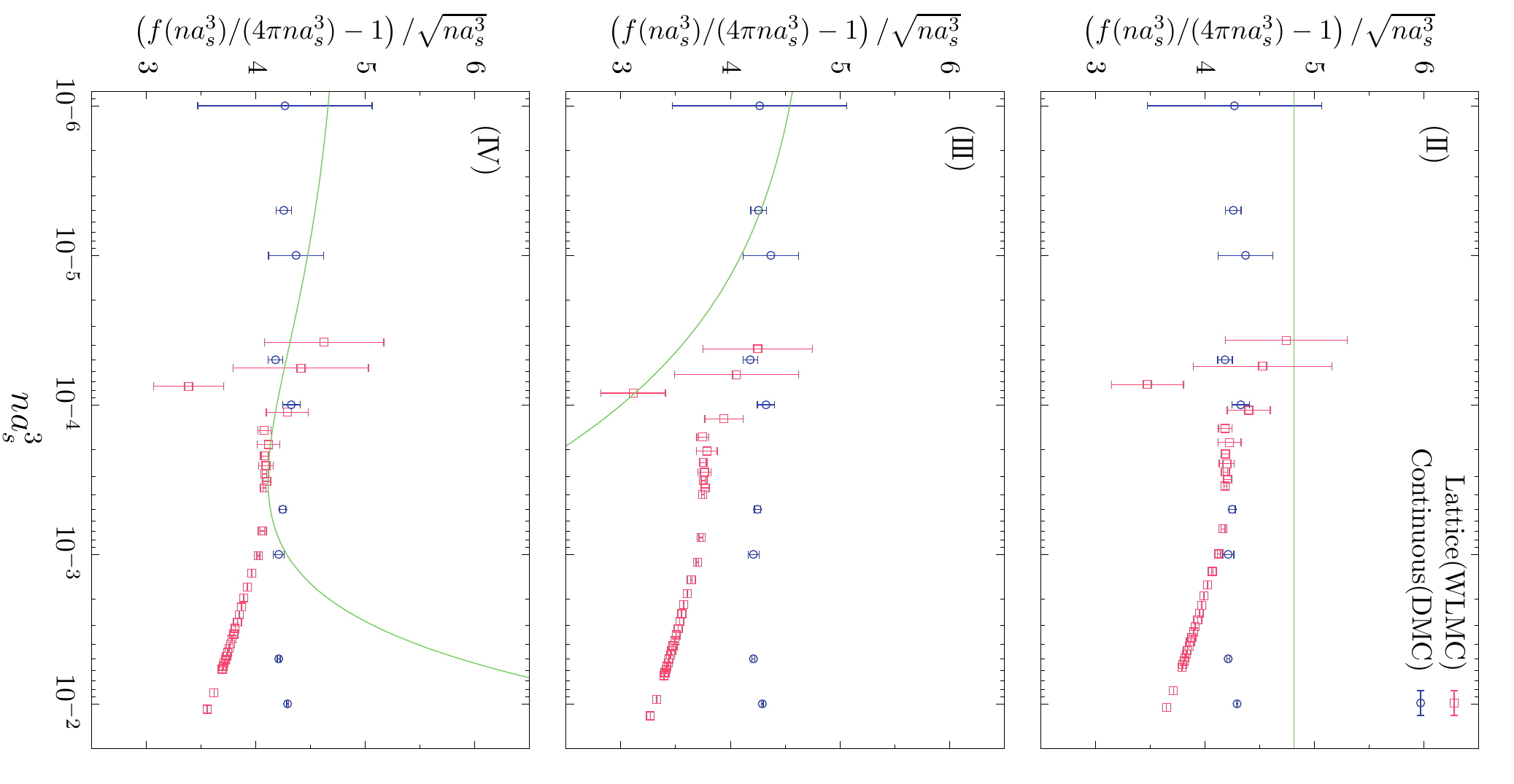}
  \end{center}
  \caption{
    (Color online)
    More detailed comparison of the numerical and the analytical
    result for the energy.
      The interpolated canonical ensemble averages are shown.
    Panels (II), (III) and (IV)
    corresponds to the fitting functions \Rom{2}, \Rom{3}, and \Rom{4},
    respectively in Table I.
    The data for the continuous space calculation are
    taken from the work by Giorgini et al.\cite{Giorgini}.
  }
  \label{fig:wnII}
\end{figure}
  In order to examine the effect of each correction term more closely, we show fittings corresponding to the three sets of parameters listed as \Rom{2}, \Rom{3} and \Rom{4} in Table~\ref{table:c}. The three panels in Figure \ref{fig:wnII} show the grand canonical data from the Monte Carlo simulation with the first term subtracted and normalized so that the second term is represented by a constant, i.e.,
  \begin{equation}
    g(x) \equiv \frac{f(x)/(4\pi x) - 1}{\sqrt{x}}
  \end{equation}
  Note that we fit only the data of the lattice system, while the behavior agrees well with the continuous-space system in the dilute region.
  The LHY analytical prediction for the coefficient
  is used in (\Rom{2}), as represented by the horizontal straight line. Even though the results of the lattice system and the continuous system are close to each other in the region $na_s^3 < 10^{-3}$ with the best estimate of $\gamma$,
  the functional form in \Rom{2} does not represent the numerical data well. In the case of \Rom{3}, which includes up to the logarithmic correction term, the lattice system and continuous system agree within the statistical errors in the region $na_s^3 < 10^{-4}$. However, the analytical results (fitting curve) rapidly deviate above $na_s^3 \sim 10^{-5}$. In the case of \Rom{4}, the fitting function includes the potential-specific non-universal coefficient $c_3$. Using the value of $\gamma$ obtained from the fitting with this non-universal correction term, the lattice system can be fitted by the analytical curve up to $na_s^3 \sim 10^{-3}$.
  In this region, both the lattice and the continuous systems agree with each other by 10\% difference, though this may be outside the universal region.

  These observations suggest that while the universal analytic expansion, $f^{\rm (\Rom{3})}$, explains the energy in the extreme dilute region $na^3_s < 10^{-5}$, the non-universal correction is required for explaining the behavior of the lattice system
  up to $na^3_s \sim 10^{-3}$.
\end{document}